\documentstyle[12pt,psfig,draft,lscape,array]{nature-ejb}

\def\simlt{\mathrel{\hbox{\rlap{\hbox{\lower4pt\hbox{$\sim$}}}\hbox{$<$}}}}
\def\simgt{\mathrel{\hbox{\rlap{\hbox{\lower4pt\hbox{$\sim$}}}\hbox{$>$}}}}

\def\ale{\mathrel{\hbox{\rlap{\hbox{\lower4pt\hbox{$\sim$}}}\hbox{$<$}}}}
\def\age{\mathrel{\hbox{\rlap{\hbox{\lower4pt\hbox{$\sim$}}}\hbox{$>$}}}}

\newcommand{\chandra}{\textit{Chandra}}
\newcommand{\hst}{\textit{HST}}
\newcommand{\swift}{\textit{Swift}}

\newcommand{\lsim }{{\lower0.8ex\hbox{$\buildrel <\over\sim$}}}
\newcommand{\gsim }{{\lower0.8ex\hbox{$\buildrel >\over\sim$}}}

\def\spose#1{\hbox to 0pt{#1\hss}}
\def\arcsec{$\,^{\prime\prime}$~}
\def\Rsun{$R_\odot$}

\def\Msun{$M_\odot$}

\def\pc3{pc$^{-3}$~}
\def\cm3{cm$^{-3}$~}
\def\L*{L$_*$~}

\begin{document}

\title{\Large \bf Short $\gamma$-ray bursts from binary neutron 
star mergers in globular clusters} 

\author{
Jonathan  Grindlay\affiliation[1]
  {Harvard-Smithsonian Center for Astrophysics, 60 Garden St., 
  Cambridge, MA 02138 USA},
Simon  Portegies Zwart\affiliation[2]
  {Astronomical Institute Anton Pannekoek and Section Computational 
Science Kruislaan \\ ~~403, 1098 SJ Amsterdam, the Netherlands}, 
Stephen  McMillan\affiliation[3]
  {Department of Physics, Drexel University, 3141 Chestnut St., 
  Philadelphia, PA 19104 USA}
}

\date{\today}{}
\headertitle{Short GRBs from neutron star mergers in globular clusters}
\mainauthor{Grindlay}

\summary{The first locations of short 
gamma-ray bursts (GRBs) in elliptical galaxies suggest they 
are produced by the mergers of double neutron star (DNS) 
binaries in old stellar populations. Globular clusters, where the 
extreme densities of very old stars in cluster cores create and 
exchange compact binaries efficiently, are a natural environment to 
produce merging NSs. They also allow some short GRBs to be 
offset from their host galaxies, as opposed to DNS systems 
formed from massive binary stars which appear to remain in galactic 
disks. Starting with a simple scaling from the first  
DNS observed in a galactic globular, which will produce a short GRB in 
$\sim$300My, we present numerical simulations which show that 
$\sim$10--30\% of short GRBs may be produced in globular clusters vs.  
the much more numerous DNS mergers and short GRBs predicted for galactic 
disks.  Reconciling the rates suggests the disk short GRBs are more beamed, 
perhaps by both the increased merger angular momentum from the DNS 
spin-orbit alignment (random for the DNS systems in globulars) 
and a larger magnetic field on the secondary NS.}  

\maketitle

\noindent{\it This paper has been submitted to Nature Physics. In accordance
with the editorial policy of Nature, it is embargoed for discussion 
with the press. Any questions should be directed to Josh Grindlay, 
jgrindlay@cfa.harvard.edu.}

\medskip

The origin of cosmic gamma-ray bursts, like most astronomical 
mysteries, required precise source positions and identification 
at other wavelengths for the first breakthroughs.  
After nearly 25 years, the so-called ``long'' GRBs 
(\gsim2-200sec), were located by coded aperture imaging of 
their hard X-ray emission which enabled more precise 
positions from their soft X-ray 
afterglows\cite{cfh+97} and then optical counterparts 
\cite{pgg+97} to be found. The optical 
identifications established them to be 
at cosmological distances. They are now understood to be due to 
relativistic jets produced in the core collapse of massive stars 
to form stellar mass black holes in certain "hyper" supernova 
events\cite{mac99}. However, about a third of GRBs were previously 
recognized\cite{kmf+93} to be distinctly different: both shorter 
duration (\lsim0.2-2 sec) and harder spectra. The 
recently-launched \swift~ satellite\cite{gehr04} was designed 
to solve the origin of the short GRBs by rapid detection of their 
X-ray and/or optical afterglows to enable their identification. 

The first of these short GRBs, GRB050509b, was 
detected and located precisely (\lsim5\arcsec) 
enough by \swift\cite{gehr05} 
to enable its plausible optical identification\cite{bloom05} with 
the halo of an elliptical galaxy at 1.12 Gpc distance. A second 
short burst, GRB0507024, was located precisely by 
\swift\cite{gcn3667,gcn3678} with an optical counterpart detected  
at an offset of $\sim$2.5kpc from the center 
of another elliptical (E2) galaxy at very 
similar redshift (z = 0.257) and hence distance\cite{berg05}. 
Since elliptical galaxies have long ceased active star formation, 
this suggested\cite{berg05} that short GRBs are associated with old 
stellar populations and likely due to the merger of two neutron 
stars or a neutron star (NS) and stellar black hole (BH) in a 
compact binary system as originally suggested for GRBs 
generally\cite{eichler89}. 
The offset position for GRB050509b at 
40$\pm$13kpc from its elliptical galaxy G1\cite{bloom05} would 
previously be attributed\cite{bloom99} to the ejection of 
the progenitor compact binary from 
the galaxy by the kick(s) imparted to it by the supernova 
event(s) which created its constituent neutron star(s) or black hole. 
This implies a kick velocity \lsim600 km s$^{-1}$ to remain bound 
to the elliptical, and yet \gsim200 km s$^{-1}$ to have its 
apogalacticon in the halo. However, this is inconsistent 
with it being like the Hulse-Taylor binary pulsar\cite{huls75} 
or the six other DNS systems now 
known in our Galaxy that were produced from the evolution of a massive 
binary system\cite{bhat91,pzy98} since recent analysis\cite{dewi05} 
shows these are all likely low-kick velocity (\lsim50 km s$^{-1}$) 
systems and thus are expected to remain within the central potential 
of their parent galaxies. Thus short GRBs from 
spiral or star formation galaxies, where massive binary evolution 
is producing DNS systems, are expected in their disks.  
Since the third short GRB located by 
\swift\cite{gladgcn05} is also associated with an elliptical, 
an alternative origin for their DNS progenitors is suggested.

We suggest instead that short GRBs are at least partly produced by 
DNS systems which formed in globular cores by exchange interactions 
of NSs into previously formed compact binaries  
containing a NS and low mass secondary. Subsequent mergers of 
these dynamically formed DNS compact binaries produced the 
short GRBs located by \swift. There is 
direct evidence that DNS systems form by dynamical interactions 
in globular clusters: the millisecond pulsar M15-C 
with a NS binary companion in the core collapsed globular cluster 
M15 in our Galaxy\cite{and90} must have formed  within its 
$\sim$100 My spindown age and then was scattered 
out of the cluster core by a NS exchange interaction\cite{phin91} 
with a cluster low mass X-ray binary (LMXB) to produce a DNS 
with present eccentricity e = 0.68 and binary period 
P = 0.33d. We demonstrate this formation scenario, 
and find that significantly more 
DNS systems are produced by NSs in core collapsed globular 
clusters (like M15) by the exchange of a NS 
for the lower mass secondaries in quiescent 
low mass X-ray binary (qLMXB) or millisecond pulsar 
(MSP) binary systems in the cluster cores. A previous 
scenario\cite{hansen98} for long GRBs from 
globular clusters, brought to our 
attention after final submission of this paper, invoked 
NSs ``smothered'' in stellar 
merger collisions to form BHs (for which cross sections remain 
unknown) but produced comparable rates to our detailed 
results for re-exchange production of DNS systems. 

Recent \chandra~ and \hst~ observations\cite{gri02,bogd05} have 
shown that MSPs in globular cluster cores do indeed undergo 
subsequent encounters to re-exchange their evolved low mass 
secondaries for a main sequence (MS) star, which is not found 
or expected\cite{bhat91} for isolated evolution from a 
LMXB or qLMXB. The 30ms spin period of the M15-C pulsar 
implies it did not complete its spinup (90\% of MSPs in globulars 
have shorter periods\cite{cam05}), so that its original companion, 
as a LMXB, was thus a $\sim$0.4--0.7\Msun~ MS star (most probable 
in the core) that was exchanged for a cluster NS, in many cases another 
MSP. The re-exchange formation model developed here can operate with 
a wide range of original secondaries (most often $\sim$0.05--0.2\Msun~ 
He white dwarfs). 

Given the existence of M15-C, can the globular cluster populations 
of elliptical galaxies plausibly contain enough DNS systems to account 
for a significant fraction of the short GRB rate? The answer is yes. 
We start with the conservative assumption that each \L* galaxy 
contains a globular cluster system with 10$^3$ globulars (typical 
for \L* ellipticals) and that the number density of such galaxies is 
$\sim$0.01 Mpc$^{-3}$ as determined\cite{cnb+01} from 2MASS and 2dF 
galaxy counts, yielding a globular cluster space 
density n$_{GC} \sim$ 10Mpc$^{-3}$. Then, using the ``local rate'' of 
observed short GRBs derived by Schmidt\cite{schmidt01} as 
R$_{GRB} \sim$0.08 Gpc$^{-3}$ yr$^{-1}$, the ``observed'' 
fractional number per globular of DNS systems 
like M15-C, with its $\tau_{merge} \sim$300My merger 
time, would need be f$_{obs}$ = R$_{GRB} \tau_{merge}$/n$_{GC}   
\sim$2.4 x 10$^{-3}$ to account for 
short GRBs. Correcting for the short GRB 
beaming factor of $\sim$30-50 inferred\cite{berg05} for (only!) 
one short GRB, GRB050724, multiplies the observed value  
for the total number of short GRB progenitors. However this 
is partly offset by the pulsar beaming factor for MSPs\cite{kramer98}, 
B$_{MSP}$ \gsim2, where the lower limit applies if DNS systems contain 
at least one MSP. Thus the net beaming factor B$_{DNS}$ = B$_{GRB}$/B$_{MSP}$ 
\gsim15 and the total number of DNS systems needed per globular is 
f$_{DNS}$ = f$_{obs}$ B$_{DNS}$ \lsim0.04. Thus short 
GRBs  could then arise 
from a population of DNS systems in globular clusters 
around \L* galaxies if they were produced within $\sim$4\% of 
its globulars. If, as we show below, this largely arises 
in the $\sim$20\% of globulars that have undergone core 
collapse\cite{tdk93}, then DNS systems are required in 
$\sim$20\% of the post core collapse (PCC) clusters, which 
would suggest another $\sim$7 of the $\sim$40 PCC globulars in the 
Galaxy should contain a M15-C-like system. This is consistent 
with currently incomplete globular cluster surveys for pulsars which 
already find 8 MSPs in binaries with large eccentricities\cite{cam05},  
many of which could have NS companions. 

Motivated by this simple scaling, we have conducted numerical
scattering experiments using the {\tt scatter3} and {\tt sigma3}
programs\cite{mcmhut96} in the {\tt Starlab} software
environment\cite{pzwart01} to compute cross sections for the formation
of DNS systems with merger times less than the age of the universe, as
well as the distributions of their emergent velocities, retention
probabilities, final periods and eccentricities.  Each experiment
consists of a three-body scattering between a target binary containing
a 1.4{\Msun} NS primary and a MS or helium WD secondary, with
parameters chosen to represent the MSP, qLMXB and (relatively rare)
LMXB targets that a lone NS in the cluster might encounter.  Our
target outcome is an exchange interaction leading to the formation of
a DNS binary with a merger time smaller than a typical cluster age,
$\sim10$ Gyr.

We adopt the following parameters for our experiments: MS secondaries
are assumed to have masses of 0.2, 0.4, or 0.8\Msun, while the WD
masses are 0.05 or 0.2\Msun.  The initial binary orbits have periods
ranging from 0.01 to 100d (as appropriate for WD or MS secondaries)
and zero eccentricity, as observed (very nearly) for most MSPs (and as
expected for their progenitor qLMXBs).  Incoming NS velocities are 10
km s$^{-1}$, a typical cluster velocity dispersion.  We include the
possibility of stellar collisions during the interaction.  The
secondary radii are the zero-age MS radii for the MS stars, and 0.03
and 0.018\,{\Rsun} for the 0.05 and 0.2{\Msun} WD secondaries,
respectively.  NS radii are taken to be 15\,km.  We performed a
total of $\sim3 \times 10^6$ simulations, leading to statistical
errors in our derived cross sections of \lsim3\%.

\begin{figure}
\centerline{\psfig{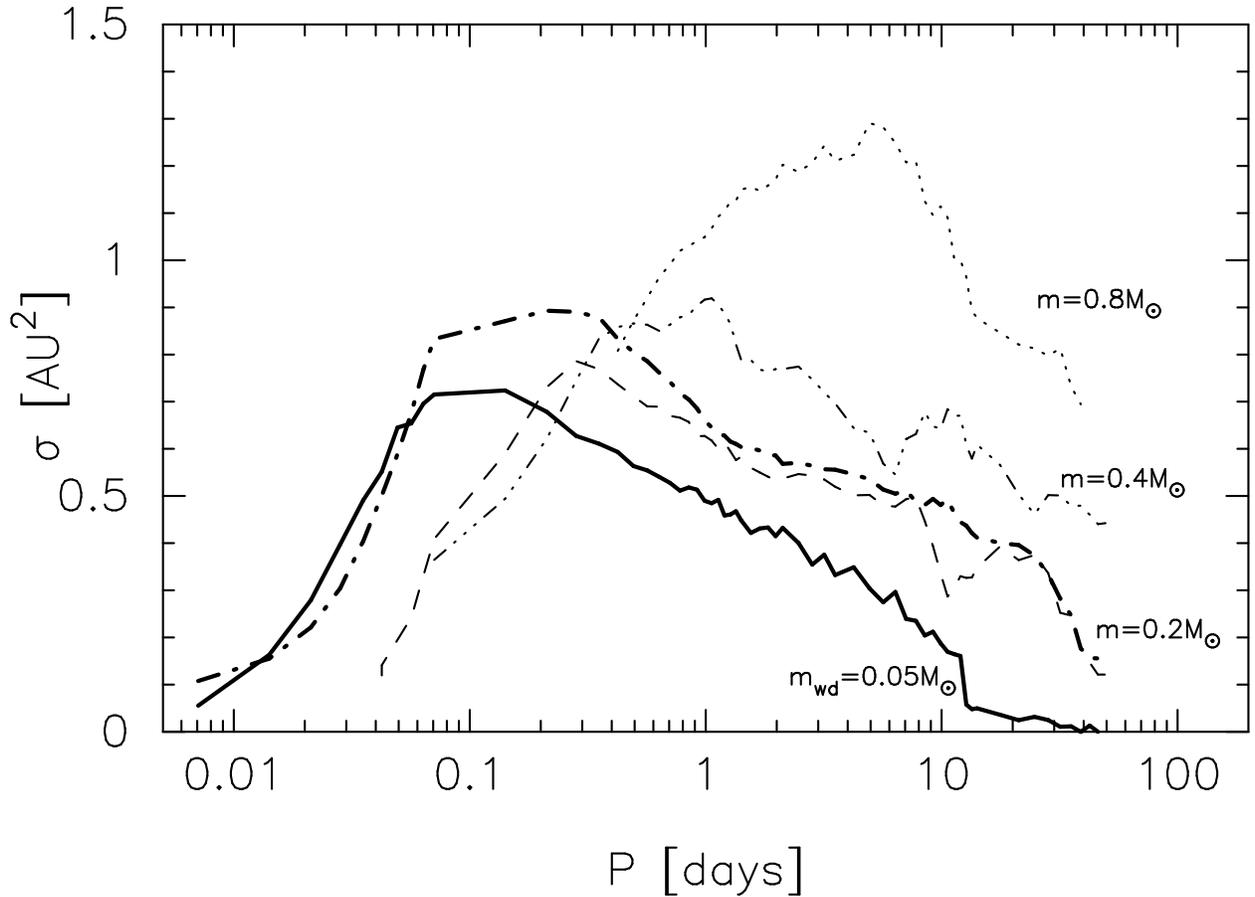}}
\caption[]{{\bf Cross sections for DNS production by exchange of  
low mass secondaries (in MSPs or LMXBs) 
with mass indicated for a NS.} The 0.05\Msun~ and 
0.2\Msun~ secondaries are calculated for He-WD mass-radius models 
and shown as the heavy curves,  whereas the remaining \gsim0.2\Msun~ 
cases are for main sequence stars. Target binary periods span 
the range shown.  Corresponding compact binaries 
represented as actual ``targets" are: LMXBs, with MS secondaries and 
masses \gsim0.2 \Msun; qLMXBs, with (typically) 0.05\Msun~ or  
0.2\Msun WD~~ or 0.2\Msun~ main sequence 
secondaries; and MSPs, with 0.05\Msun~ or  
0.2\Msun~ WD secondaries. The cross sections are comparable 
($\sim$0.8 AU$^2$) over the distribution of target 
orbital periods (mostly $\sim$0.1--1d) of MSPs\cite{cam05} or 
LMXBs in globulars. Total DNS numbers were derived from these 
cross sections, NS densities and number of progenitor binary 
systems for clusters near core collapse (see text).} 
\label{fig:xsec}
\end{figure}

\begin{figure}
\centerline{\psfig{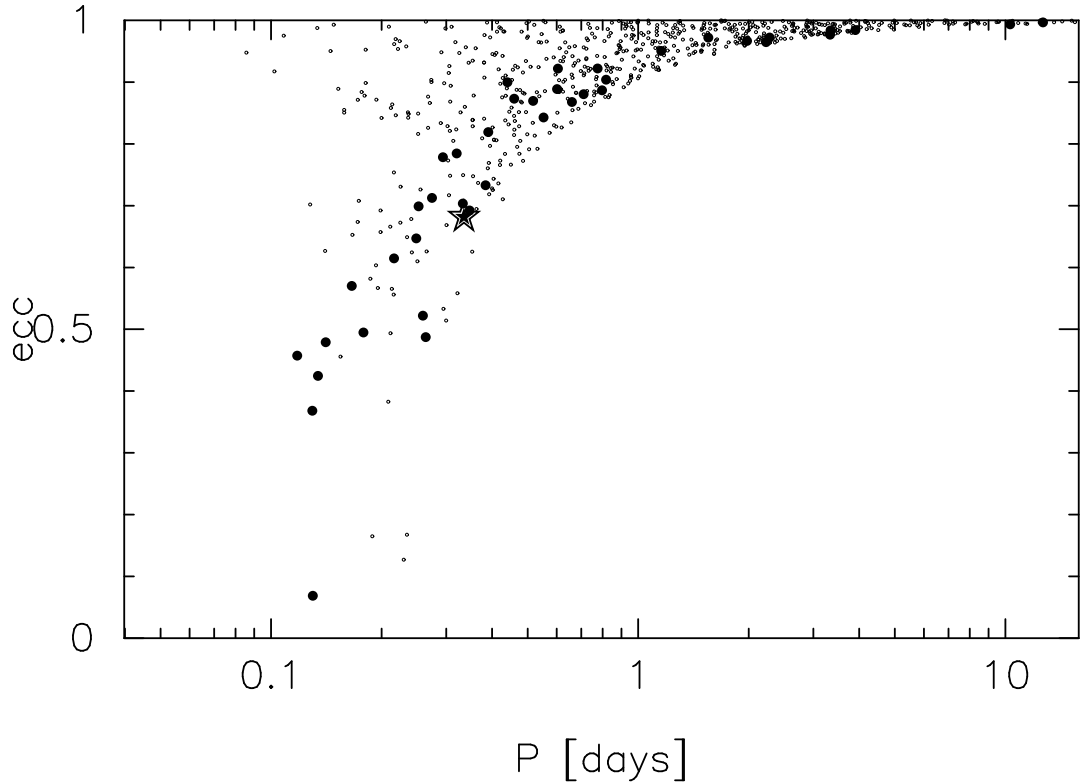}}
\caption[]{{\bf Binary period vs. eccentricity for DNS systems 
formed by NS exchange into LMXBs with 0.4\Msun~ secondaries.} 
Only DNS binaries produced with merger times \lsim10 Gy are plotted. 
Each ``heavy''  point marks a binary 
that has survived until today but will merge within a 
Hubble time, whereas the initially created DNS 
systems are marked by the small dots (most of which have merged). 
The {\bf Star} marks the parameters 
for the M15-C system, showing that it indeed could be produced by 
an exchange encounter of a field NS with an LMXB (or qLMXB) containing 
a NS and 0.4\Msun~ secondary.}
\label{fig:PvsEcc}
\end{figure}

The results are quite remarkable: M15-C systems could be produced with 
cross sections that are comparable ($\sim$0.8 AU$^2$) over target 
binary periods P$_b \sim$0.1--1d (Figure~\ref{fig:xsec}). Even 
more striking is the range of output DNS periods (P$_b$) and 
eccentricities (e) produced by these exchange encounters 
(Figure~\ref{fig:PvsEcc}). For the 0.4\Msun~ MS secondary 
case appropriate to M15-C, and the exchange interaction randomly 
distributed between $t=0$ and
$t=10^{10}$\,years,  the orbital evolution due to the emission of
gravitational wave radiation is subsequently followed for each of
the DNS systems produced until the current age of the Galaxy of
$10^{10}$\,years. The star indicates the observed position of the binary
pulsar M15-C and is ``predicted'' by the distribution of 
survivors (heavy dots). 

With the cross sections in Fig~\ref{fig:xsec} we can estimate the
number of DNS systems produced per globular.  We assume that the
orbital period distribution of progenitor systems is similar to that
observed for MSPs in Galactic globular clusters\cite{cam05}, resulting
in an average cross section of $\sigma \simeq 0.8$\,AU$^2$. We note
that this greatly exceeds the area of the target binary orbit due to
gravitational focusing at low encounter velocity ($\sim$10 km
s$^{-1}$).  The instantaneous formation rate of DNS binaries which
will merge within 10 Gyr is then given by $\Gamma = N_{\rm pr} n
\sigma v$, where $N_{\rm pr}$ is the number of potential progenitor
systems in a cluster with mean neutron star number density $n$ and
velocity dispersion $v$.  After substitution and scaling to parameters
typical of globular clusters, we obtain
\begin{equation}
	\Gamma_{\rm DNS} \simeq 4
	                 \left(n \over [10^6\,{\rm pc}^{-3}]\right)
	                 \left(\sigma \over [0.8\,{\rm AU}^2]\right)
			 \left(v \over [10\,{\rm km s}^{-1}]\right) 
			 \left(N_{\rm pr} \over [20]\right)
			 \ {\rm Gyr}^{-1}\, {\rm globular}^{-1}\, ,
\label{Eq:rate}
\end{equation}
where, as discussed below, we normalize to a
neutron star density of $n\sim 10^6$\,pc$^{-3}$ and a total of $N_{\rm
pr} = 20$\, progenitor binaries containing a NS primary.  The latter
figure is estimated from the population of target LMXBs ($\sim$0.05),
qLMXBs ($\sim$1) and MSPs ($\sim$10) per cluster, as derived from
{\chandra} studies of (primarily) non-core-collapse but high 
central density clusters\cite{heinke05}, 
but increased by a factor of 2 for the expected production during 
core collapse of additional target LMXBs, MSP-WD and MSP-MS systems. 

The quantity $\Gamma_{DNS}$ varies dramatically over a cluster's
evolutionary lifetime, rising sharply around the time of core
collapse.  Assuming a 10\% neutron-star retention
fraction\cite{pfahl02}, we estimate
that the Galaxy's ``non-core-collapse'' clusters have
$n\sim10^2-10^4\,{\rm pc}^{-3}$, while models for core-collapse 
clusters\cite{dull97} predict $n\gg10^6\,{\rm pc}^{-3}$. The 
cluster central velocity dispersion $v$
depends only weakly on the state of the core\cite{dull97}, 
scaling as $n^{0.05}$. As a result, during the
homologous late stages of core collapse the central density $n$ scales
roughly as $\tau^{-1.2}$, where $\tau = t_{cc}-t$ is the time
remaining until collapse, and the total number of DNS systems produced
in the cluster, $N_{DNS}=\int\Gamma_{DNS}\,dt$, is dominated by the
behavior near $\tau=0$.  Thus post-core-collapse (PCC) 
clusters, $\sim$20\% of the total in the Galaxy\cite{tdk93}, 
would dominate the overall production of DNS binaries in globulars.

We use the M15 simulation data of Dull et al.\cite{dull97} 
to resolve the formal singularity at $\tau=0$, and find that
\begin{equation}
	\int_{t_{cc}-\tau_0}^0\,n(t)\,dt ~\approx~ 20 n_0 \tau_0\,,
\label{Eq:integral}
\end{equation}
where $\tau_0$ is the time remaining until core collapse 
from a state with central density 
$n_0 = n(t_{cc}-\tau_0)$.  The dependence of $n$ on $\tau$ makes
the precise instant at which this expression is evaluated relatively
unimportant.  We choose $n_0 = 10^6\,{\rm pc}^{-3}$,
corresponding to $\tau_0\sim100 t_{rc}\sim100{\rm Myr}$ for parameters
appropriate to M15.  
The behavior of clusters during the PCC 
phase remains poorly understood but simulations indicate that 
the reexpansion is significantly slower than the collapse, and may 
include a series of nonlinear core oscillations. 
Thus we expect that the post-collapse phase will contribute at least
as much to $N_{DNS}$ as did the collapse; each recollapse will
contribute roughly as much again. Weighting the cross sections of the
DNS binaries with the observed orbital period distribution for MSPs in
globulars,\cite{cam05} since they dominate the DNS formation, we then 
estimate the total number of merging DNS binaries {\em per 
PCC cluster} as
\begin{equation}
	N_{\rm DNS} \simeq 12
	             \left(\sigma \over [0.8{\rm AU}^2]\right)
		     \left(v \over [10 {\rm km s}^{-1}]\right) 
		     \left(N_{\rm pr} \over [20]\right)
		     \left(\tau_0 \over [100\,{\rm Myr}]\right)\,.
\label{Eq:number}
\end{equation}
The coefficient 12 in Eq.~(\ref{Eq:number}) comes from $\sim25$\% of
the progenitors scattering directly into merging orbits, and a further
$\sim35$\% forming wider DNS systems that harden into merging orbits
following a subsequent interaction during the core-collapse phase. 
We note that $\sim$10\% of these systems will stem from (q)LMXBs 
and hence should be comparable to M15-C.  The remainder,
descended from MSPs, will be wider and more eccentric.

Thus, combining the two populations (qLMXBs/LMXBs and MSPs), ignoring
the small differences in donor mass in qLMXBs and LMXBs, we derive a 
total of $\sim$12 DNS binaries per core-collapse cluster, or $\sim 480$ 
DNS systems in the Galactic system of $\sim$40 PCC globulars formed over 
the age of the Galaxy. Some 400 of these have already merged due to 
the emission of gravitational waves (we estimate that in the last 1 
Gyr $\sim$40 systems have merged), and $\sim$80 survive today (but 
will merge within a Hubble time, T$_h$). The majority of DNS systems
are expected to remain in the globular cluster, as the exchange
encounters tend to induce only a small recoil velocity on the binary
pulsar. Only the LMXBs have significant ($\sim$40\%) ejection of DNS
pulsars into the Galactic disk and halo, as suggested for
M15-C\cite{phin91}.  Our derived total of $\sim$80 DNS systems
surviving in the globular cluster system of the Galaxy 
(i.e., $\sim$2 per PCC globular) that will merge
within T$_h$ and produce short GRBs is $\sim10\times$ larger than our
simple scaling rate from M15-C alone. The difference comes from
the fact that M15-C itself has a shorter than average merger time; our
calculations now sample the entire distribution of merger times.

The derived DNS merger rate in the Galaxy of 40 per Gyr in 200 (total) 
globular clusters implies, for the same
scalings to globular clusters in external galaxies,  $\sim$2 DNS
mergers per year per Gpc$^3$. With the estimated short 
GRB beaming factor\cite{berg05} B$_{GRB}$, this gives  
an observable GRB rate of $\sim$0.07 Gpc$^{-3}$ yr$^{-1}$, or roughly
the Schmidt rate\cite{schmidt01} but a factor of $\sim$1.5--100 
below the rates (with large uncertainties) recently obtained by Guetta
and Piran,\cite{gp05} for 
DNS production tied to the star formation rate. 
Given the large uncertainties in short GRB beaming 
factors, short GRB rates, and the details of core collapse 
(e.g., core collapse oscillations could increase our rates 
by factors \gsim2), and the additional DNS production from 
cluster primordial binaries containing NSs (most NSs retained 
in the cluster are likely in binaries), 
we conclude that perhaps $\sim$10--30\% of short GRBs are from DNS 
mergers in globulars.  

Our DNS merger rate in globulars in the Galaxy of $\sim$400 
in 10$^{10}$ years or $\sim$0.04\,Myr$^{-1}$ yields a 
detection rate for ground based gravitational wave
antennas that is \gsim200$\times$ smaller 
than the rates summarized\cite{pzy98,gp05} for DNS mergers in high mass 
binaries in the Galactic disk.
Either these high mass progenitor rates for DNS production in the disk 
have been significantly overestimated, or most do not 
produce short GRBs, or their beaming angles 
are much smaller. We note that the NS population in globulars 
is likely dominated by recycled pulsars with low magnetic 
fields (B $\sim10^{8-9}$ Gauss) whereas in the disk DNS systems   
the second-born NS, with no recycling, will have B $\sim10^{12-14}$ 
Gauss. Although this B field is inconsiquential for 
the merger, it may launch a magnetically driven jet 
with smaller beaming angles. Combined with 
the randomly aligned spin-orbital angular momenta of 
DNS systems in globulars vs. the aligned spin-orbit 
vectors for disk systems, disk short GRBs might have beaming angles 
\gsim10$\times$ smaller to reconcile the rates. 

The recent discovery of an interstellar medium (ISM) with 
density n$_{ISM} \sim$ 0.1 cm$^{-3}$ in the 
globular cluster 47 Tucanae\cite{freire01} from the variable dispersion 
measure of its radio-detected MSPs is consistent with the ISM 
density deduced\cite{berg05} for GRB050724. Globular clusters that have 
just traversed their host galaxy disks will have lower ISM densities, 
so that fainter afterglows are expected for some short GRBs, such as 
appears to be the case (given the non-detection) 
for GRB050509b\cite{gehr05}.

Finally, the short GRB050709 discovered with HETE\cite{graz05} 
was followed by a faint, soft and delayed 
($\sim$150s) X-ray flare which may point 
to a different origin. Mergers of BH-NS binaries may 
disrupt the NS and produce a ``delayed'' short GRB, 
whereas the lack of direct evidence for stellar mass BHs in 
globulars restricts the sample to NS-NS mergers. 
Indeed GRB050709 is reported\cite{berg05} to be 
associated with a star formation galaxy and so is 
more plausibly a DNS merger from a massive binary although 
of course its host galaxy also presumably includes globular clusters. 
Thus short GRBs plausibly arise from two populations, as suggested on other 
grounds\cite{gp05}: DNS or NS-BH mergers from high mass systems 
in star-forming disks with smaller beaming and DNS mergers 
from dynamically formed systems in globulars. The latter may be 
confirmed by further identifications with ellipticals or locations 
in host galaxy halos; the globular clusters themselves are too 
faint for detection with magnitudes V \gsim29 even at z $\sim$0.1.

\begin{acknowledge} 
JG thanks Neil Gehrels for an initial discussion of the first 
short GRB observed with \swift.  This work was 
supported in part for JG by NASA grant 
NNG04GK33G and for SM by NASA grant NNG04GL50G. SPZ is 
supported in part by the Royal Netherlands Academy 
of Arts and Sciences (KNAW) and the Netherlands Advanced School for Astronomy 
(NOVA). The authors thank Ed van den Heuvel for comments.
\end{acknowledge}

\medskip

{\small Correspondence and requests  should be addressed to 
JG (josh@cfa.harvard.edu).}

\clearpage

\clearpage

\end{document}